\documentclass[12pt]{article}
\usepackage{graphicx}
\usepackage{cmap}
\usepackage{xcolor}

\usepackage{color}
\usepackage{hyperref}

\usepackage[T2A]{fontenc}
\usepackage[utf8]{inputenc}
\usepackage[english]{babel}
\usepackage{amsmath}
\usepackage{amssymb}
\usepackage{amsthm}
\usepackage{mathrsfs}

\newlength{\wdth}

\newif\ifru
\rufalse

\newif\ifen
\entrue

\newtheorem{enTheorem}{Theorem}
\newtheorem{enProposition}{Proposition}
\newtheorem{enDefinition}{Definition}
\newtheorem{enLemma}{Lemma}
\newtheorem{enRemark}{Remark}

\begin{document}  

\title{On Bellman equation in the limit order optimization problem for high-frequency trading}

\author{M.I.~Balakaeva\footnote{M.V. Lomonosov Moscow State University \& Institute for Information Transmission Problems (A.A. Kharkevich Institute), Moscow, Russian Federation}, \; \& \; 
A.Yu.~Veretennikov\footnote{
Institute for Information Transmission Problems (A.A. Kharkevich Institute) \& S.M. Nikolskii Mathematical Institute, RUDN university, Moscow, Russian Federation}
}

\maketitle

\ifen
\begin{abstract}
An approximation method for construction of optimal  strategies in the 
bid \& ask limit order book in the high-frequency trading (HFT) is studied. The basis is the article  \cite{5}, in which certain seemingly serious gaps have been found; in the present paper they are carefully corrected. However, a bit surprisingly, our corrections do not change the main answer in \cite{5}, so that, in fact, the gaps turn out to be unimportant. An explanation of this effect is offered.
\end{abstract}
\fi

\ifen
\section{Introduction}
It is assumed that there are no transaction costs in the action market, and that there is no remuneration to be paid. The market is assumed to be complete.  The Bachelier model is accepted as the average action price $S_u$,  also called mid-price, that is, this price satisfies the equation  
\begin{equation}\label{S_t}
dS_u=\sigma dW_u \quad (S^{t, s}_T= s + \sigma (W_T - W_t)), 
\end{equation}
with the initial value $S_t=s$ at time $t$, where $W_u$ is the standard Wiener process, and $\sigma>0$ is a constant. This section is a squeezed  presentation of the setting from \cite{5}. The main part of the present note is section \ref{sec:main}, which is a gently corrected version of the well-known presentation in \cite{5}. We highlight that dispite the amendments, the main approximate result in the theorem \ref{thm1} in what follows confirms the conclusions in \cite{5}. 

~

The agent's goal is to maximize the expected utility function at the expiry time  $T$. The latter function may or may not be linear; here we accept an exponential function\footnote{Presumably, the ``real'' utility function could be $1-\exp(-\gamma (x+qs))$; however, this $+1$ does not affect in any way the calculus, which is to follow.} 
\[
 -\exp(-\gamma (x+qs)), 
\]
where $x$ is the initial capital in dollars, and $s$ stands for the current action price. Respectively, the agent's value function with the horizon $T$ is 
\[
v(x,s,q,t)= \mathsf{E}_t [ -\exp(-\gamma (x+qS_T)) ], 
\]

So, the task is to determine the reservation bid and ask prices for this model (notice that the latter is not yet fully determined, though). Given the equation (\ref{S_t}), the function $v$ may be rewritten in the form,

\begin{equation}\label{func_v}
  v(x,s,q,t)=-\exp(-\gamma x)\exp(-\gamma q s)\exp(\frac{\gamma^2q^2\sigma^2(T-t)}{2})
\end{equation}


Let us determine the reservation bid and ask prices for the agent. They must satisfy the property of the {\bf indifference} after buying or selling one share of stock. 
    
\begin{enDefinition}\label{def1}
The reservation bid price $r^b$ for the value function $v$ is determined from the equation 
\[v(x-r^b(s,q,t),s,q+1,t)=v(x,s,q,t); 
\]  
the reservation ask price $r^a$ is determined from the equation
    \[v(x+r^a(s,q,t),s,q-1,t)=v(x,s,q,t)\]    
\end{enDefinition}

\begin{enLemma}\label{st1}
Reservation prices $r^a$ and $r^b$ equal, respectively, 
\begin{equation}\label{r_ab}
\begin{split}
r^a(s,q,t)=s+(1-2q)\frac{\gamma \sigma^2(T-t)}{2},\\\\ 
      r^b(s,q,t)=s+(-1-2q)\frac{\gamma \sigma^2(T-t)}{2}.\\
\end{split} 
\end{equation}   
\end{enLemma}
For the proof see Appendix.

\fi

\ifen

\subsection{Optimization over the infinite horizon}
Intuitively, for the HFT, any finite fixed time $T>0$ is a ``very large'' value, given that each ``tick'' is extremely small. Hence, let us introduce an artificial ``stationary version'' of our expected utility as follows. 

\begin{enDefinition}\label{def3}
Expected ``stationary'' utility function\footnote{This is an intuitive term; actually, we do not assume that the process $S_t$ is stationary in a strong or weak sense.} over the infinite horizon is given by the formula,
\[ \bar{v}=\mathsf{E}\biggl[\int_0^\infty -\exp(-wt)\exp(-\gamma(x+qS_t))dt\biggl].
\]
\end{enDefinition}
Here $w>0$ is one more parameter, which has the meaning of a discount. By using Gaussian integration, 
we obtain the formula as stated in the next lemma.

\begin{enLemma}\label{st2}

\begin{equation}
\bar{v}
= \frac{2\exp(-\gamma x)\exp(-\gamma qs)}{\gamma^2 q^2 \sigma^2-2w}.
\end{equation}
\end{enLemma}

Similarly to the previous point, we can now also define {\em ``stationary''} versions of the reservation bid and ask prices.
\begin{enDefinition}\label{def4}
``Stationary'' reservation bid and ask prices are determined by the following equations: 
\[
\bar{v}(x+\bar{r}^a(s,q,t),s,q-1,t)=\bar{v}(x,s,q,t),
\]
and
\[\bar{v}(x-\bar{r}^b(s,q,t),s,q+1,t)=\bar{v}(x,s,q,t).
\]    
\end{enDefinition}

\begin{enLemma}\label{st3}
``Stationary'' reservation bid and ask prices  $\bar{r}^a(s,q)$ and $\bar{r}^b(s,q)$ are given by the following explicit expressions, respectively, 
\[
\bar{r}^a(s,q)=s+\frac{1}{\gamma}\ln\left(1+\frac{(1-2q)\gamma^2 \sigma^2}{2w-\gamma^2 q^2 \sigma^2}\right),
\]
and
\[
\bar{r}^b(s,q)=s+\frac{1}{\gamma}\ln\left(1+\frac{(-1-2q)\gamma^2 \sigma^2}{2w-\gamma^2 q^2 \sigma^2}\right),
\]
where \ $w>\frac12 \gamma^2\sigma^2q^2$.
\end{enLemma}
The proof follows from trivial computation with Gaussian integrals.
Thus, the parameter $w$ defines the upper limit of the inventory position that our agent is allowed to occupy.

\fi

\ifen

\subsection{Limit orders}
Now let us consider an agent who may buy and sell actions via establishing his limit orders around the mid-price    \eqref{S_t}. Namely, he determines his bid price $p^b$ and ask price  $p^a$ and would be obliged to buy or, respectively, to sell actions if the mid-price attains the corresponding levels. The following new notations will be useful.

\begin{enDefinition}\label{def5}
Let $s$ be the action mid-price. Then
\begin{equation}\label{db}
\delta^b:=s-p^b,
\end{equation}
and
\begin{equation}\label{da}
\delta^a=p^a-s.
\end{equation}
\end{enDefinition}
To finish the setting, it is necessary to discuss one more crucial notion of this model: the schedule of execution of the limit orders. Of course, the agent does not buy or sell all available actions on the market whenever the orders can be realised, but (in this model) he does it with some rate. The latter is modeled by two independent Poisson processes $N^a_t$ and $N^b_t$, with intensities  $\lambda^a$ and $\lambda^b$, respectively. These intensities will be shortly discussed; at the moment we consider them to be known functions depending both on the values $\delta^a$ and $\delta^b$, respectively. 
The execution of limit orders of the agent by market orders (that is, which have no fixed prices and which are executed according ot the best available prices immediately) will be modelled by independent Poisson processes with intensities depending on the distance between the limit (bid and ask) price by the agent and the current market prices. Namely: 

\begin{itemize}
\item{Limit ask orders:}
\begin{itemize}
\item The agent places an ask  (sell) order for the price \( p^a = s + \delta^a \), where:

\begin{itemize}
            
\item \( s \) is a current market price  (for example, a mid-price),
\item \( \delta^a \geq 0 \) is the distance from \( s \) to the order price.
\end{itemize}

\item The intensity of execution \( \lambda^a(\delta^a) \) is a decreasing function of \( \delta^a \).
 \\
\textit{In our case:} \( \lambda^a(\delta^a) = A \exp(-\kappa \delta^a) \), where \( A, \kappa > 0 \).
\item Market \textbf{buy} order  executes the agent's ask limit order with the intensity \( \lambda^a(\delta^a) \).
\end{itemize}

\item{Limit bid (buy) orders:}
\begin{itemize}
\item The agent places the bid (buy) order for the price \( p^b = s - \delta^b \), where \( \delta^b \geq 0 \).
\item The execution intensity \( \lambda^b(\delta^b) \) is a decreasing function of  \( \delta^b \).
\\
\textit{In our case: } \( \lambda^b(\delta^b) = A \exp(-\kappa \delta^b) \).
\item Market \textbf{sell} orders execute the agent's limit bid order with the intensity  \( \lambda^b(\delta^b) \).
\end{itemize}
\end{itemize}

Intuitively, the further the agent's quotes are from the market mid-price, the less frequently he/she will get market limit orders for buy and sell. 
\\
In this model the evolution of the agent's portfolio is stochastic and depends on the market limit orders for buy and sell. Indeed, his/her capital (respectively, number of stocks) increase every time when the buy order (resp., sell order) is executed.
 \\
This may be written in the form, 
\[
dX_t=p^a dN_t^a-p^b dN_t^b,
\]
where $N^b_t$ is how many actions the agent bought, and $N^a_t$ is the number of actions sold. As it was  said earlier, here 
$N^b_t$ and $N^a_t$ are Poisson processes with intensities $\lambda^b$ and $\lambda^a$, respectively. The number of actions in the agent's portfolio at time $t$ then equals 
\[
q_t = N^b_t - N^a_t,  
\]
assuming that $q_0=0$. 

The agent's goal is a maximization of a value function, 
\begin{equation}\label{maxfl}
u(s,x,q,t)=
\max_{\delta^a {\ge 0},\delta^b {\ge 0}} \mathsf{E}_t\left[ -\exp(-\gamma(X_T+q_T S_T)) \right].
\end{equation}
It may be noticed that in this situation, unlike in the case of the previous (sub)section, the agent may control the bid and ask prices, and, hence, implicitly affects the flow (flows) of the orders he/she obtains.

One of the main area of the econophysics society 
was the description of the market microstructure laws. In particular, the intensity function $\lambda(\delta)$ introduced earlier was studied in the papers 
~\cite{1,2}. One possible option for this function, accepted also in \cite{5}, turned out to be 
\[ 
\lambda (\delta) = A\exp(-\kappa \delta), \; \text{where} \;
 A=\frac{\Lambda}{\alpha} \text{ and } \kappa=\alpha K.
 \]

\fi

\ifen

\section{Main results}\label{sec:main}
Recall that the agent's goal (from the mathematical point of view) is to solve a  maximization problem (\ref{maxfl}) by choosing a {\bf strategy}\footnote{Here, of course, the infinite upper limit is an exaggeration; however, we follow all the simplifications of the model  accepted in \cite{5}.} $(\delta^a, \delta^b) \in [0,\infty)^2$.  
This type of problems/models was  studied in \cite{3} by Ho and Stoll, with a reference to \cite{Davis}. 
One of the key steps in their analysis was the use of the dynamic programming principle and establishing Bellman's equation for the function $u$:
 
\begin{equation}\label{BEu}
\begin{cases}
\begin{aligned}   u_t+\frac12\sigma^2 u_{ss}
&+\max_{\delta^b {\ge 0}}\lambda^b (\delta^b )\left[ u(s,x-s+\delta^b,q+1,t)
-u(s,x,q,t)\right]  \\ 
&+ \max_{\delta^a{\ge 0}}\lambda^a (\delta^a)\left[ u(s,x+s+\delta^a,q-1,t)
-u(s,x,q,t)\right]=0, \\      
\end{aligned}
 \\
u(s,x,q,T)=-\exp(-\gamma(x+qs)).  
\end{cases}
\end{equation}
By introducing a substitution 
\[
u(s,x,q,t)=-\exp(-\gamma x)\exp(-\gamma \theta (s,q,t)),  
\]
this equation (\ref{BEu}) may be rewritten in terms of the new unknown function to be determined, $\theta (s,q,t))$, as stated in the next proposition.  

\begin{enProposition}\label{st3}

The equation (\ref{BEu}) is equivalent to the equation on the function   $\theta (s,q,t)$,
\begin{equation}\label{BEt}
\begin{cases}
\begin{aligned}   \theta_t+\frac12\sigma^2 \gamma \theta^2_{ss}-\frac12 \sigma^2 \gamma \theta^2_s
&+\max_{\delta^b\ge 0} \left[\frac{\lambda^b(\delta^b)}{\gamma}\left[       1-e^{\gamma(s-\delta^b-r^b)}\right] \right]  \\ 
&+\max_{\delta^a\ge 0} \left[\frac{\lambda^a(\delta^a)}{\gamma}\left[ 
1-e^{-\gamma(s+\delta^a-r^a)}\right] \right]=0,  
 \\    
\end{aligned}
 \\
\theta(s,q,T) = qs.
\end{cases}
\end{equation}
\end{enProposition}
By using the relations  
\begin{equation}\label{eq_ln1}
s-r^b(s,q,t)=\delta^b-\frac{1}{\gamma} \ln
\left( 1-\gamma \frac{\lambda^b(\delta^b)}{\frac{\partial \lambda^b}{\partial\delta}
(\delta^b)}\right),
\end{equation}
and     
\begin{equation}\label{eq_ln2}
r^a(s,q,t)-s=\delta^a-\frac{1}{\gamma} \ln
\left( 1-\gamma \frac{\lambda^a(\delta^a)}{\frac{\partial \lambda^a}{\partial\delta}  (\delta^a)}\right),
\end{equation}
(detailed derivation of these relationships can be seen in the Appendix ~\ref{app:spread_derivation}). 
Now, substituting  
\[ 
\lambda (\delta) = A\exp(-\kappa \delta), 
\]
we rewrite,   
\[
\!e^{\gamma(s-\delta^b-r^b)}\!=\!
e^{-\ln\left(\!1-\gamma \frac{\lambda^b(\delta^b)}{\frac{\partial \lambda^b}{\partial\delta}(\delta^b)}\!\right)}  
\!=\! \frac{\frac{\partial \lambda^b}{\partial\delta}(\delta^b)}{\frac{\partial \lambda^b}{\partial\delta}(\delta^b)-\gamma\lambda^b(\delta^b)}
\!=\!\frac{-\kappa A e^{-k\delta^b}}{-\kappa A e^{-k\delta^b}-\gamma A e^{-k\delta^b}}
\!=\!\frac{k}{k+\gamma}.
\]
Therefore, we have, 
$$ \frac{\lambda^b(\delta^b)}{\gamma}\left[ 
1-e^{\gamma(s-\delta^b-r^b)}\right]  = \frac{Ae^{-k\delta^b}}{\gamma}\left[ 
1-\frac{k}{k+\gamma}
\right]  = \frac{A}{k+\gamma} e^{-k\delta^b}  .
$$

Similarly, 
$$
\frac{\lambda^a(\delta^a)}{\gamma}\left[ 
1-e^{\gamma(s+\delta^a-r^a)}\right]  = \frac{A}{k+\gamma}
e^{-k\delta^a}.
$$

~

Thus, the system (\ref{BEt}) may be further rewritten as in the next proposition. 
\begin{enProposition}\label{st3a}
The equation (\ref{BEt}) is equivalent to 
\begin{equation}\label{simp_eq}
\begin{cases}
\theta_t+\frac12\sigma^2 \gamma \theta^2_{ss} -\frac12 \sigma^2 \gamma \theta^2_s
+ \frac{2A}{k+\gamma}(e^{-k\delta^a} + e^{-k\delta^b})=0,  
 \\ \\
\theta(s,q,T)=qs.
\end{cases}
\end{equation}
\end{enProposition}
Next, the idea is to solve the problem approximately using  asymptotic expansion in powers of $q$ (for this technique, see, for example, \cite{12}).
First, we expand both exponents in the expression with the maximum in a Taylor series, keeping only the zero and first terms in it. We have,
\begin{equation}\label{max_tey}
\begin{split}
\frac{A}{k+\gamma}\left[e^{-k\delta^b} + e^{-k\delta^a}\right] & = \frac{A}{k+\gamma}\left[2-\kappa(\delta^a+\delta^b)+\ldots \right] \\
& = \frac{2A}{k+\gamma} +\bar{o}(\delta^a + \delta^b), \quad (\delta^a, \delta^b )\to (0,0)
\end{split}
\end{equation}

\begin{enRemark}\label{rem1}
Here we suggest that the value $(\delta^a+\delta^b)$ is small, so we drop it. This is, of course, not rigorous, but is certainly in line with the idea of HFT, which is the approach allowing to gain using rather tiny spreads.  Notice that, as a result, the coefficient $\theta^{(0)}$ will not depend on $s$ (see proposition 3); otherwise, the term $\theta^{(0)}_s$ may not vanish in the equations (19) and (20), on $\theta^{(1)}$ and $\theta^{(2)}$, respectively.
\end{enRemark}

~

Next, we expand the function $\theta(s,q,t)$ into an asymptotic series\footnote{Recall that, in any case, $q$ is an integer, positive or negative, and, in any case, we are not talking here about Taylor's formula, but rather about an asymptotic series.} in powers of $q$:
\begin{equation}\label{as_exp}    \theta(s,q,t)=\theta^{(0)}(s,t)+q \theta^{(1)}(s,t)+\frac12 q^2 \theta^{(2)}(s,t)+\ldots
\end{equation}

We identified an incorrect formula in the article \cite{5}, where \( r^a(s,q,t) \) and \( r^b(s,q,t) \) were expressed erroneously. Using the expansion above and relations \eqref{eq_5} - \eqref{eq_6} (see Appendix for details), we derive:

\begin{align}
r^b(s,q,t) &= \theta(s,q+1,t) - \theta(s,q,t) = \theta^{(1)}(s,t) + \tfrac{1}{2} \theta^{(2)}(s,t)(2q+1) + \cdots \label{eq_rb} \\
r^a(s,q,t) &= \theta(s,q,t) - \theta(s,q-1,t) = \theta^{(1)}(s,t) + \tfrac{1}{2} \theta^{(2)}(s,t)(2q-1) + \cdots \label{eq_ra}
\end{align}

Using equations \eqref{eq_rb} and \eqref{eq_ra}
as well as the optimality conditions \eqref{eq_ln1} and \eqref{eq_ln2},
we find that the optimal pricing strategy is reduced to forming a spread:

\begin{equation}\label{spread_ab}
\delta^a+\delta^b=-\theta^{(2)}(s,t)+\frac{2}{\gamma} \ln\left[1+\frac{\gamma}{k}\right].
\end{equation} 
Also, we obtain, 
\[ 
r(s,q,t) =\frac{r^a+r^b}{2}=\theta^{(1)}(s,t)+\theta^{(2)}(s,t) q.
\]
The found exact expressions for the spread and average price $r(s,q,t)$ are new, due to an error in the article, noted in the previous relations. Yet, the asymptotic answers will be the same as in \cite{5}, see the theorem \ref{thm1} in what follows.

Here
$\theta^{(1)}$ can be interpreted as the reservation price when inventory is zero.
$\theta^{(2)}$ can be interpreted as the sensitivity of market maker quotes to inventory changes. For example, if $\theta^{(2)}$ is large,
accumulating a long position $q > 0$ will result in low quotes.
The bid-ask spread in \eqref{spread_ab} is independent of inventory.
This follows  from our assumption of an exponential rate of arrivals.
The spread consists of two components:
one depends on the sensitivity to inventory changes $\theta^{(2)}$,
the other on the order intake rate via the parameter $\kappa$.

\begin{enProposition}\label{prot0}
By grouping the terms at $q^0$ we obtain a system on
$\theta^{(0)}$:
\begin{equation}\label{eq_q0}
\begin{cases}
\theta^{(0)}_t+\frac12\sigma^2 \theta^{(0)}_{ss} -
\frac12 \gamma \sigma^2 (\theta^{(0)}_s)^2 
+\frac{2A}{\kappa+\gamma}
=0,  
 \\ \\
\theta^{(0)}(s,T)=0
\end{cases}
\end{equation}
The solution to the equation (\ref{eq_q0}) is
$\theta^{(0)}=\frac{2A}{\kappa + \gamma}(T-t)$.
\end{enProposition}  
\vspace{2em}
\begin{enProposition}\label{prot12}
We note that the linear term does not depend on the stock q.
Therefore, if we substitute \eqref{as_exp} and \eqref{max_tey} into \eqref{simp_eq}
and group the terms of order $q$, we obtain the following system on $\theta^{(1)}$ :

\begin{equation}\label{eq_q1}
\begin{cases}
\theta^{(1)}_t+\frac12\sigma^2  \theta^{(1)}_{ss} -
\gamma \sigma^2 \theta^{(0)}_s \theta^{(1)}_s
=0,  
\\ \\
\theta^{(1)}(s,T)=s
\end{cases}
\end{equation}
Moreover, 
by grouping the terms at $q^2$ we obtain a system on
$\theta^{(2)}$:

\begin{equation}\label{eq_q2}
\begin{cases}
\theta^{(2)}_t+\frac12\sigma^2 \theta^{(2)}_{ss} -
\gamma \sigma^2 (\theta^{(1)}_s)^2 
-\gamma \sigma^2 \theta^{(0)}_s \theta^{(2)}_s
=0,  \\ 
\\
\theta^{(2)}(s,T)=0
\end{cases}
\end{equation}
Solution of the equation
    \begin{equation*}\label{eq_q1}
      \begin{cases}
          \theta^{(1)}_t+\frac12\sigma^2 \theta^{(1)}_{ss} -
            \gamma \sigma^2 \theta^{(0)}_s \theta^{(1)}_s 
          =0,  \\           
        \\
        \theta^{(1)}(s,T)=s
      \end{cases}
\end{equation*}
can be written out explicitly: $\theta^{(1)} = s$.

Also, the solution of the equation
 \begin{equation*}\label{eq_q2}
      \begin{cases}
          \theta^{(2)}_t+\frac12\sigma^2 \theta^{(2)}_{ss} -
            \gamma \sigma^2 (\theta^{(1)}_s)^2 - \gamma \sigma^2 \theta^{(0)}_s \theta^{(2)}_s 
          =0,  \\           
        \\
        \theta^{(2)}(s,T)=0
      \end{cases}
\end{equation*}
we can write it out explicitly: $\theta^{(2)} = -\gamma \sigma^2 (T-t) $.

\end{enProposition}
Note that in the article ~\cite{5} the authors obtain other systems of equations in which they assume that
$\theta^{(0)}=0$, this does not follow from the obtained relations, and also does not follow from the definition of
the function $\theta(s,q,t)$, therefore these equations are rewritten in a different, more complex form.

\begin{enTheorem}\label{thm1}
Under the assumptions above, there are explicit formulae for the quotes and spread:

$$
r(s,q,t) = \theta^{(1)}(s,t) + \theta^{(2)}(s,t)q = s - \gamma \sigma^2 (T-t)q,  
$$
and
$$\delta^a + \delta^b = - \theta^{(2)}(s,t) + \frac{2}{\gamma}\ln\left[ 1+ \frac{\gamma}{\kappa}\right] = \gamma \sigma^2 (T-t) + \frac{2}{\gamma} \ln\left[ 1+ \frac{\gamma}{\kappa}\right].
$$
\end{enTheorem}
We emphasize that the obtained (approximate) result
coincides with the formula from the work~\cite{5}, although we previously noted and corrected some errors in their calculations. Also, it is worth mentioning that the result might be different if some presumably small quantities $\delta^a, \delta^b$ in \ref{max_tey} were not dropped; see the remark \ref{rem1}.

All proofs are in the appendix.

\fi

~


~ \hfill {\em A P P E N D I X}

{\bf Proof of lemma \ref{st1}} \\ 
$\bigtriangledown$ 

Using the (first) equality  in the definition \ref{def1} along with the representation (\ref{func_v}), we obtain,  
\begin{equation}
\begin{split}
-\exp(-\gamma (x+r^a(s,q,t)))\exp(-\gamma(q-1)s)\exp(\frac{\gamma^2 (q-1)^2\sigma^2(T-t)}{2})=
 \\
=-\exp(-\gamma x)\exp(-\gamma q s)\exp(\frac{\gamma^2 q^2 \sigma^2(T-t)}{2})
 \\
\exp(-\gamma r^a(s,q,t))\exp(\gamma s)\exp(\frac{\gamma^2 (-2q+1)\sigma^2(T-t)}{2})=1
 \\
-\gamma r^a(s,q,t)+\gamma s+(\frac{\gamma^2 (-2q+1)\sigma^2(T-t)}{2})=0
 \\  
\Rightarrow r^a(s,q,t)=s+(1-2q)\frac{\gamma \sigma^2 (T-t)}{2}.
\end{split} 
\end{equation}
Similarly, for $r^b(s,q,t)$ we get, 
\begin{equation}r^b(s,q,t)=s+(-1-2q)\frac{\gamma \sigma^2 (T-t)}{2}.
\end{equation} 
The lemma \ref{st1} is proved. 
\hfill $\bigtriangleup$\\

{\bf Proof of lemma \ref{st2}} \\ 
$\bigtriangledown$

\begin{equation}
    \begin{split}
      \bar{v} &= \mathsf{E}\biggl[\int_0^\infty -\exp(-wt)\exp(-\gamma(x+qS_t))dt\biggl] \\
      &= -\exp(-\gamma x)\exp(-\gamma qs) \int_0^\infty \exp(-wt)\mathsf{E}[\exp(-\gamma q \sigma W_t)]dt \\
      &= -\exp(-\gamma x)\exp(-\gamma qs) \int_0^\infty \exp(-wt)\exp\left(\frac{\gamma^2 q^2 \sigma^2 t}{2}\right)dt \\
      &= -\exp(-\gamma x)\exp(-\gamma qs) \int_0^\infty \exp\left(t\left(\frac{\gamma^2 q^2 \sigma^2}{2}-w\right)\right)dt \\
      &= \frac{2\exp(-\gamma x)\exp(-\gamma qs)}{\gamma^2 q^2 \sigma^2-2w} 
    \end{split} 
\end{equation}

Let's consider for $\bar{r}^b(s,q)$ \ ($\bar{r}^a(s,q)$ is derived similarly):
   
    \begin{align}
      \bar{v}(x-\bar{r}^b(s,q,t),s,q+1,t) &= \bar{v}(x,s,q,t)\notag\\
      \frac{2\exp(-\gamma (x-\bar{r}^b(s,q))\exp(-s\gamma (q+1)))}{\gamma^2 (q+1)^2\sigma^2-2w}  
      &= \frac{2\exp(-\gamma x)\exp(-\gamma q s)}{\gamma^2 q^2 \sigma^2-2w}\notag\\
      \exp(-\gamma x)\exp(\gamma \bar{r}^b(s,q)) 
      &=\frac{(\gamma^2 (q+1)^2 \sigma^2 - 2w)(\exp(-\gamma x)\exp(-\gamma q s))}{(\gamma^2 q^2 \sigma^2 - 2w)\exp(-s\gamma(q+1))}\notag\\
      \exp(\gamma \bar{r}^b(s,q))
      &=\frac{(\gamma^2 (q+1)^2 \sigma^2 - 2w)\exp(s\gamma)}{(\gamma^2 q^2 \sigma^2 - 2w)}\notag\\
      \gamma \bar{r}^b(s,q)
      &=\ln\left(1+\frac{\gamma^2 (2q+1)\sigma^2}{\gamma^2 q^2 \sigma^2-2w}\right)+s\gamma\notag\\
      \bar{r}^b(s,q)
      &=\frac{1}{\gamma}\ln\left(1+\frac{\gamma^2 (2q+1)\sigma^2}{\gamma^2 q^2 \sigma^2-2w}\right)+s \notag
    \end{align} 
The lemma \ref{st2} is proved.
\hfill $\bigtriangleup$\\

~

{\bf Proof of optimal bid and ask spread (equations \eqref{eq_ln1}, \eqref{eq_ln2})}

The relations \eqref{eq_ln1} and \eqref{eq_ln2} are derived from the first-order conditions of the optimization problems for market makers. We focus on the bid side; the derivation for the ask side is analogous.

~

\noindent{Bid Side Optimization:}
Market makers solve the problem:
\[
\max_{\delta^b \geq 0} \ f(\delta^b), \quad \text{where} \quad f(\delta^b) = \frac{\lambda^b(\delta^b)}{\gamma} \left( 1 - e^{\gamma(s - \delta^b - r^b)} \right).
\]
We assume the arrival rate $\lambda^b(\delta^b)$ is a positive, strictly decreasing, and differentiable function of the spread $\delta^b$.

\noindent{First-Order Condition:}
The first derivative is given by:
\begin{align*}
\frac{\partial f}{\partial \delta^b}
&= \frac{1}{\gamma} \left[
\frac{\partial \lambda^b}{\partial \delta^b} \left( 1 - e^{\gamma(s - \delta^b - r^b)} \right)
+ \lambda^b(\delta^b) \cdot \gamma e^{\gamma(s - \delta^b - r^b)}
\right].
\end{align*}
Setting the derivative to zero, $\partial f/\partial \delta^b = 0$, yields:

\begin{align*}
\frac{\partial f}{\partial \delta^b} 
&= \frac{1}{\gamma} \left[ 
\frac{\partial \lambda^b}{\partial \delta^b} \left( 1 - e^{\gamma(s - \delta^b - r^b)} \right) 
+ \lambda^b(\delta^b) \cdot \gamma e^{\gamma(s - \delta^b - r^b)} 
\right] = 0
\end{align*}
Simplifying the expression:
\[
\frac{\partial \lambda^b}{\partial \delta^b} \left( 1 - e^{\gamma(s - \delta^b - r^b)} \right) 
= - \gamma \lambda^b(\delta^b) e^{\gamma(s - \delta^b - r^b)}
\]
Multiply both sides by $e^{-\gamma(s - \delta^b - r^b)}$:
\[
\frac{\partial \lambda^b}{\partial \delta^b} \left( e^{-\gamma(s - \delta^b - r^b)} - 1 \right) 
= - \gamma \lambda^b(\delta^b)
\]
Solving for the exponential term:
\[
e^{-\gamma(s - \delta^b - r^b)} = 1 - \gamma \frac{\lambda^b(\delta^b)}{\partial \lambda^b/\partial \delta^b}
\]

\[
-\gamma(s - \delta^b - r^b) = \ln \left( 1 - \gamma \frac{\lambda^b(\delta^b)}{\partial \lambda^b/\partial \delta^b} \right)
\]
We obtain the final expression:
\[
s - r^b = \delta^b - \frac{1}{\gamma} \ln \left( 1 - \gamma \frac{\lambda^b(\delta^b)}{\partial \lambda^b/\partial \delta^b} \right) =
\delta^b - \frac{1}{\gamma} \ln \left( 1 + \frac\gamma\kappa \right)
\]
which corresponds to equation \eqref{eq_ln1}.

The function $f(\delta^b)$ is continuous on $[0, +\infty)$ with $\lim\limits_{\delta^b \to +\infty} f(\delta^b) = 0$. Under the assumption that $f(\delta^b)$ is strictly concave and has exactly one critical point on $(0, +\infty)$, this point must be the global maximizer.

For the specific case $\lambda^b(\delta^b) = A e^{-\kappa \delta^b}$, the second derivative can be computed explicitly:
\[
f''(\delta^b) = \frac{A}{\gamma} e^{-\kappa \delta^b} \left[ \kappa^2 - (\kappa + \gamma)^2 e^{\gamma(s - \delta^b - r^b)} \right].
\]
At the critical point $\delta^{b*}$, the second derivative is negative, confirming the concavity and thus the maximum.

~

\noindent\textbf{Optimization for Ask Side:}

The market maker's problem for the ask side is:
\[
\max_{\delta^a \geq 0} \ f(\delta^a), \quad \text{where} \quad f(\delta^a) = \frac{\lambda^a(\delta^a)}{\gamma} \left( 1 - e^{-\gamma(s + \delta^a - r^a)} \right).
\]

The first derivative is:
\[
\frac{\partial f}{\partial \delta^a} = \frac{A}{\gamma} e^{-\kappa \delta^a} \left[ -\kappa + (\kappa + \gamma) e^{-\gamma(s + \delta^a - r^a)} \right].
\]
Setting the derivative to zero yields the optimal spread:
\[
\delta^{a*} = r^a - s + \frac{1}{\gamma} \ln \left( 1 - \gamma \frac{\lambda^a(\delta^a)}{\partial \lambda^a/\partial \delta^a} \right).
\]

The same reasoning applies to the optimization problem for $\delta^a$: the function $f(\delta^a)$ is continuous on $[0, +\infty)$ with $\lim\limits_{\delta^a \to +\infty} f(\delta^a) = 0$, and under the concavity assumption with a unique critical point, this point must be the global maximizer.

For $\lambda^a(\delta^a) = A e^{-\kappa \delta^a}$, the second derivative is:
\[
f''(\delta^a) = \frac{A}{\gamma} e^{-\kappa \delta^a} \left[ \kappa^2 - (\kappa + \gamma)^2 e^{-\gamma(s + \delta^a - r^a)} \right],
\]
which is negative at the critical point $\delta^{a*}$, confirming the maximum.

~

{\bf Derivation of \( r^a \) and \( r^b \)}\\
Using asymptotic series:
\begin{align*}
\theta(q+1,s,t) &= \theta^{(0)} + (q+1)\theta^{(1)} + \tfrac{1}{2}(q+1)^2\theta^{(2)} + \cdots \\
\theta(s,q,t) &= \theta^{(0)} + q\theta^{(1)} + \tfrac{1}{2}q^2\theta^{(2)} + \cdots \\
\Rightarrow r^b(s,q,t) &= \theta^{(1)} + \tfrac{1}{2}\theta^{(2)}(2q+1) + \cdots \tag{\ref{eq_rb} reproduced}
\end{align*}

\begin{align*}
\theta(s,q-1,t) &= \theta^{(0)} + (q-1)\theta^{(1)} + \tfrac{1}{2}(q-1)^2\theta^{(2)} + \cdots \\
\Rightarrow r^a(s,q,t) &= \theta^{(1)} + \tfrac{1}{2}\theta^{(2)}(2q-1) + \cdots \tag{\ref{eq_ra} reproduced}
\end{align*}

{\bf Spread Derivation}\\
From optimality conditions:
\begin{multline*}
\delta^a + \delta^b = r^a - r^b + \frac{1}{\gamma} \ln\left[ \left(1 - \gamma \frac{A e^{-\kappa \delta^b}}{-\kappa A e^{-\kappa \delta^b}}\right) \left(1 - \gamma \frac{A e^{-\kappa \delta^a}}{-\kappa A e^{-\kappa \delta^a}}\right) \right] \\
= \underbrace{\left[ \theta^{(1)} + \tfrac{1}{2}\theta^{(2)}(2q-1) \right] - \left[ \theta^{(1)} + \tfrac{1}{2}\theta^{(2)}(2q+1) \right]}_{r^a - r^b = -\theta^{(2)}} + \frac{2}{\gamma} \ln\left(1 + \frac{\gamma}{\kappa}\right)
\end{multline*}
yielding \eqref{spread_ab}.

~

{\bf Proof of proposition \ref{st3}}

$\bigtriangledown$  
 \\
  \indent The solution to this nonlinear equation is continuous in the variables $s$, $x$, and $t$
and depends on discrete values of the endowment $q$.
Due to our choice of exponential utility, we can simplify the problem
by the following substitution (now the new unknown is $\theta(s,q,t)$):

  \[u(s,x,q,t)=-\exp(-\gamma x)\exp(-\gamma \theta (s,q,t)) \]

 Let's make this substitution:

 \[ u_t=-\exp(-\gamma x)(-\gamma)\exp(-\gamma \theta(s,q,t))\theta_t(s,q,t);  \]
 \[ u_s=-\exp(-\gamma x)(-\gamma) \exp(-\gamma \theta(s,q,t)) \theta_s(s,q,t);  \]

 \[
  \begin{aligned}
    u_{ss} &= -\exp(-\gamma x)\gamma^2 \exp(-\gamma \theta(s,q,t))\theta^2_s(s,q,t)+\\
    &+ (-\exp(-\gamma x)(-\gamma)\exp(-\gamma \theta(s,q,t))\theta_{ss}(s,q,t))\\
\end{aligned}
\]  
\[ 
\begin{aligned}
\!u(s,x\!-\!s\!+\!\delta^b,q\!+\!1,t)\! -\! u(s,x,q,t)\! =\!
&-\!\exp(-\gamma(x\!-\!s\!+\!\delta^b))\exp(-\gamma \theta(s,q\!+\!1,t))\\
&+ \exp(-\gamma x)\exp(-\gamma \theta(s,q,t))\\
\end{aligned}
\]
\[=-\exp(\gamma x) \left[ \exp(\gamma s-\gamma \delta^b)\exp(-\gamma \theta(s,q+1,t))
-\exp(-\gamma \theta(s,q,t))\right]\]

    $  u(s,x+s+\delta^a,q-1,t)-u(s,x,q,t) = $
    \[ =- \exp(-\gamma x) \left[ \exp(-\gamma (s+\delta^a))\exp(-\gamma \theta(s,q-1,t)) \\
    - \exp(-\gamma \theta(s,q,t))\right].
\]
    \indent By the definition of the reservation prices of demand and supply \eqref{def1}
and by the assignment of the utility function (obtained above), we find that
\begin{equation}\label{eq_5}
r^b(s,q,t)=
\theta(s,q+1,t) - 
\theta(s,q,t),  
\end{equation}
\begin{equation}\label{eq_6}
r^a(s,q,t)=\theta(s,q,t)-\theta(s,q-1,t),   \end{equation}
    
Taking these relations into account and reducing the factor $-\exp(-\gamma x)$
the system is transformed to the desired form. 
~
\hfill $\bigtriangleup$

~

{\bf Derivation of equation (\ref{spread_ab})}\\

  \indent Using equations \eqref{eq_rb} and \eqref{eq_ra}
and the optimality conditions \eqref{eq_ln1} and \eqref{eq_ln2},
we find that the optimal pricing strategy is reduced to forming a spread:

\begin{align*}
&\delta^a+\delta^b=r^a(s,q,t)-r^b(s,q,t)+\frac{1}{\gamma}\ln
    \left[ (1-\gamma \frac{Ae^{-k\delta^b}}{-kAe^{-k\delta^b}})(1-\gamma \frac{Ae^{-k\delta^a}}{-kAe^{-k\delta^a}})\right]
    \\\\ 
&= \theta^{(1)}(s,t)+\frac12 \theta^{(2)}(s,t)(2q-1)+\ldots
    -(\theta^{(1)}(s,t)+\frac12 \theta^{(2)}(s,t)(2q+1)+\ldots).
\end{align*}
Hence, we have approximately,       
    \begin{equation}
      \delta^a+\delta^b=-\theta^{(2)}(s,t)+\frac{2}{\gamma} \ln\left[1+\frac{\gamma}{k}\right], 
    \end{equation}
and
\[
r(s,q,t)=\frac{r^a+r^b}{2}=\theta^{(1)}(s,t)+\theta^{(2)}(s,t) q.
\]
~

{\bf Proof of proposition \ref{prot0}}

\indent First of all, to solve the equations we need the Feynman-{Kaс } formula, which states that an equation of the form:

\begin{equation}\label{eq_FK}
\begin{cases}
u_t+\mu(x,t)u_x+ \frac12 \sigma^2(x,t) u_{xx} - V(x,t)u+f(x,t)=0  ,  \\ \\
u(x,T)=\psi(x) \notag
\end{cases}
\end{equation} 
has a solution that can be expressed in the following form:
\[
u(x,t)=\mathsf{E}^Q \left[ \int_t^T e^{-\int_0^s V(X_{\tau})d\tau} f(X_s,s)ds+e^{-\int_t^T V(X_\tau)d\tau}
\psi(X_T) \mid X_t=x \right] , 
\]
where $Q$ is a probability measure such that the random process $X_t$ is an Ito process,
described by the stochastic equation
\[
dX_t= \mu (X_t,t)dt + \sigma(X_t,t)dW^Q_t
\]

Next,

\noindent $\bigtriangledown$
\\
\indent In equation (\ref{eq_q0}),
i.e.
  $\theta^{(0)} = \frac{1}{\gamma} \ln u$. \\
  \[
  \theta_t^{(0)} = \frac{1}{\gamma}\frac{u_t}{u}, \
   \theta_s^{(0)} = \frac{1}{\gamma}\frac{u_s}{u}, \
   \theta_{ss}^{(0)} = \frac{1}{\gamma}\frac{u_{ss}u-u_s^2}{u^2}  
  \]
Let's rewrite the equation in new notations:
  \begin{equation*}\label{eq_q3}
    \begin{cases}
        u_t+\frac12 \sigma^2 u_{ss} + a u = 0\text{,  где  } a  = \frac{2A}{\kappa+\gamma}\gamma  \\ 
      \\
      u(T,s)=1
    \end{cases}
\end{equation*}
Let's use the Feynman-Kac formula with $\mu(x,t)=0$, $-V(x,t) = a$, \\
$\psi(x)=1$, $f(x,t)=0$.

Then
\begin{align*} 
u(x,t) 
&=\mathsf{E} \left[ \int_t^T e^{-\int_0^s V(X_{\tau})d\tau} f(X_s,s)ds+e^{-\int_t^T V(X_\tau)d\tau}
\psi(X_T) \mid X_t=x \right] \\
&= \mathsf{E} \left[ e^{a(T-t)}\mid X_t=x \right] 
= \mathsf{E} \left[ e^{a(T-t)} \right] = e^{a(T-t)}
  \ \ \ \ \ \ \ \ \ \ \ \ \ \ \ \ \ \ \ \  \ \ \ \ \ \ \ \  \bigtriangleup
\end{align*}

~

{\bf Proof of proposition \ref{prot12}} \\
$\bigtriangledown$ \\

Solution of the equation
\begin{equation*}\label{eq_q1}
      \begin{cases}
          \theta^{(1)}_t+\frac12\sigma^2 \theta^{(1)}_{ss} -
            \gamma \sigma^2 \theta^{(0)}_s \theta^{(1)}_s 
          =0,  \\           
        \\
        \theta^{(1)}(s,T)=s
      \end{cases}
  \end{equation*}
    we can write it out explicitly: $\theta^{(1)} = s$.

Let's rewrite the equation taking into account the calculated function $\theta^{(0)}$:
  \begin{equation*}\label{eq_q1}
      \begin{cases}
          \theta^{(1)}_t+\frac12\sigma^2 \theta^{(1)}_{ss} -
          =0,  \\           
        \\
        \theta^{(1)}(s,T)=s
      \end{cases}
  \end{equation*}
Let's use the Feynman--Kac formula with $\mu(x,t)=0$, $-V(x,t) = 0$, \\
$\psi(x)=x$, $f(x,t)=0$.

Then
  $$ 
      \theta^{(1)}(x,t) 
      = \mathsf{E} \left[ \psi(X_T) \mid X_t=x \right] \\
      = \mathsf{E} \left[X_t + \sigma (W_T - W_t) \mid X_t=x \right]  = x
  $$
$\hfill \bigtriangleup$

Solution of the equation \begin{equation*}\label{eq_q2}
      \begin{cases}
          \theta^{(2)}_t+\frac12\sigma^2 \theta^{(2)}_{ss} -
            \gamma \sigma^2 (\theta^{(1)}_s)^2 - \gamma \sigma^2 \theta^{(0)}_s \theta^{(2)}_s 
          =0,  \\           
        \\
        \theta^{(2)}(s,T)=0
      \end{cases}
  \end{equation*}
    we can write it out explicitly: $\theta^{(2)} = -\gamma \sigma^2 (T-t) $.


  \noindent $\bigtriangledown$
  \\
  \indent 
  Let's rewrite the equation taking into account the calculated function $\theta^{(0)}$ и $\theta^{(1)}$:
  \begin{equation*}\label{eq_q1}
      \begin{cases}
          \theta^{(2)}_t+\frac12\sigma^2 \theta^{(2)}_{ss} - \gamma \sigma^2
          =0,  \\           
        \\
        \theta^{(2)}(s,T)=0
      \end{cases}
  \end{equation*}
Let us use the Feynman-Kac formula with $\mu(x,t)=0$, $-V(x,t) = 0$, \\
$\psi(x)=0$, $f(x,t)= - \gamma \sigma^2$. Then

$$ 
\theta^{(2)}(x,t) 
      = \mathsf{E} \left[ - \gamma \sigma^2 (T-t)\mid X_t=x \right]   = - \gamma \sigma^2 (T-t).
  $$
$\hfill \bigtriangleup$


\end{document}